\documentclass[11pt,twoside]{article}
\usepackage{newpasp}

\index{Malhotra, S}
\index{Rhoads, J}

\begin{document}

\title{Large Area Lyman Alpha survey: finding young galaxies at z=4.5}
\author{Sangeeta Malhotra}
\affil{Johns Hopkins University}
\author{James Rhoads}
\affil{Space Telescope Science Institute}
\author{Arjun Dey, Buell Jannuzi}
\affil{Kitt Peak National Observatory}
\author{Daniel Stern, Hyron Spinrad}
\affil{ University of California,Berkeley}

\begin{abstract}

Strong Lyman-$\alpha$ emission is a signpost of young stars and the
absence of dust and thus indicates young galaxies. To find such a
population of young galaxies at z=4.5 we started the Large Area Lyman
Alpha survey (LALA). This survey achieves an unprecedented combination
of volume and sensitivity by using narrow-band filters on a large
format (36 \arcmin $\times 36 \arcmin $) camera on the 4 meter
telescope at KPNO. The volume density and star-formation contribution
of the Lyman-$\alpha$ emitters at z=4.5 is comparable to that of Lyman
break galaxies.  With many candidates and a few spectroscopic
confirmations in hand we discuss what the properties of Ly-$\alpha$
emitters imply for galaxy and star formation in the early universe.

\end{abstract}

% Include keywords if you wish. The keywords.apj file, found on aas.org 
% in the pubs/aastex-misc directory, contains a list of keywords used 
% with the ApJ and Letters.  

%%\keywords{infrared: galaxies -- galaxies: nuclei -- galaxies: starburst}

% That's it for the front matter.  On to the main body of the paper.

\section{Introduction}

More than three decades ago Partridge and Peebles (1967) predicted
that galaxies in their first throes of star-formation should be strong
emitters in the Ly-$\alpha$ line. Their predictions were
optimistic, based on converting roughly 2\% of gas into metals in $3
\times 10^7$ years in Milky Way sized galaxies, which translates into
a line luminosity of $ \approx 10^{44} erg/s$. These objects are also
expected to be common - if all the $L^*$ galaxies have undergone such
a phase of rapid star-formation one should see a surface density of
about $3 \times 10^3 $ per square-degree. Searches based on these
expectations did not detect Ly-$\alpha$ emitters (see review by
Pritchet 1994). Only recently have Ly-$\alpha$ emitters been
discovered, (Lowenthal et al 1997; Cowie \& Hu 1998; Hu, Cowie \&
McMahon 1998; Dey et al 1998; Hu, McMahon, \& Cowie 1999; Kudritzki et
al 2000, Rhoads et al. 2000), but at luminosities roughly a hundred
times fainter and the total number discovered remains small.

\section{LALA Survey}

In order to get statistically useful samples we started an efficient
search for Ly-$\alpha$ emitters (and other emission line galaxies)
in 1998 using the CCD Mosaic Camera at the Kitt Peak National
Observatory's 4m Mayall telescope. The Mosaic camera has eight $2048
\times 4096$ chips in a $4 \times 2$ array comprising a $36'\times
36'$ field of view. The final area covered by the LALA survey is
$0.72$ square-degrees in two MOSAIC fields.  Five overlapping narrow
band filters with full width at half maximum (FWHM) $\approx 80$\AA\
are used. The total redshift coverage is $4.37 < z < 4.57$.  This
translates into surveyed comoving volume of $8.5 \times 10^5$ $Mpc^3$
per field for $H_0 = 70 km s^{-1} Mpc^{-1}$, $\Omega = 0.2$,
$\Lambda=0$. About 70\% of the imaging at $z \approx 4.5$ is complete,
and an extension of the survey to $z>5$ has started. In about 6 hours
per filter per field we are able to achieve line detections of about
$2 \times 10^{-17} ergs\ cm^{-2}\ s^{-1}$. Broadband images of these
fields in a custom $B_w$ filter ($\lambda_0 = 4135$\AA,
$\hbox{FWHM}=1278 $\AA) and the Johnson-Cousins $R$, $I$, and $K$
bands are being taken as part of the NOAO Deep Wide-Field Survey
(Jannuzi \& Dey 1999).

\begin{figure}[h]
\plotfiddle{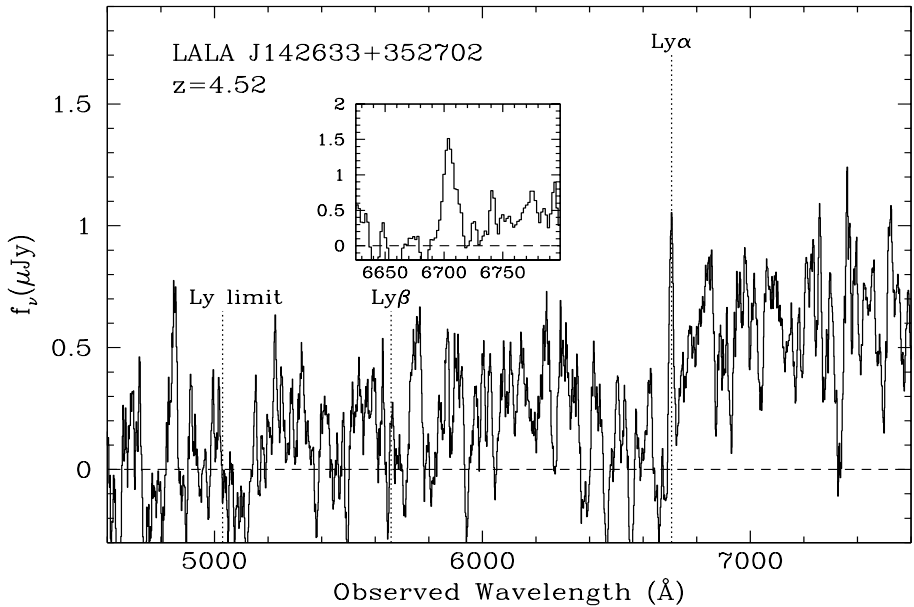}{3.5in}{-90}{47}{42}{-190}{255} 
\caption{ A confirmed $z=4.52$ Ly-$\alpha$ source.  This object has
a line flux of $1.7 \times 10^{-17} ergs\  cm^{-2}\  s^{-1}$ and observed
equivalent width of 84\AA.  The line is asymmetric and has a strong
continuum decrement from the red to the blue side of the line. Another
drop in the continuum level can be also be discerned at the Lyman
limit.}
\end{figure}

Pilot spectroscopic studies were carried out to confirm emission line
candidates. This is necessary because the narrow band selection
criterion picks up [OIII] emitters at z=0.3, [OII] emitters at z=0.8
as well as Ly-$\alpha$ emitters at z=4.5. In case of [OIII] emitters at
z=0.3, the spectra show three lines, two of [OIII] and H$\beta$ and in
most cases, H$\alpha$. We can often see the [OIII] and H$\beta$ lines
for the [OII] emitters depending on the wavelength coverage. Some of
the [OII] and all Ly-$\alpha$ emitters show only one line each in
the wavelength range covered, so a one-line spectrum may be either a
z=0.8 or a z=4.5 object.  The confirmation of the Ly-$\alpha$ line
is largely circumstantial. An asymmetric line and a continuum drop
blueward of the line due to intergalactic absorption are the
signatures of the Ly-$\alpha$ line (Stern \& Spinrad 1999). In
sources where we do not detect the continuum, the high equivalent
width of the line provides the (somewhat weaker) evidence that the
line in question is indeed Ly-$\alpha$.

Spectroscopic follow-up of a cross-section of emission line candidates
was obtained with the LRIS instrument (Oke et al.~1995) at the Keck
10m telescope on 1999 June 13 (UT) and April 2000. The June 1999 mask
yielded one, perhaps two, Ly-$\alpha$ emitters. The April 2000 mask
had a better detection rate as we had used broad band colors to reduce
the contamination by foregroud emission line galaxies. It yielded 4
more Ly-$\alpha$ emitters.

\section{Discussion}

From the few spectroscopic confirmations and many candidates we
estimate about 4000 emitters per square degree per unit redshift
(Rhoads et al. 2000). Within the present uncertainties the number
density of Ly-$\alpha$ emitters is 3 times higher than the Lyman
break galaxies (Steidel et al. 1999) of similar brightness and
redshifts.  The star-formation rates estimated for the Ly-$\alpha$
galaxies are 0.5-2 times the star-formation rate derived for continuum
selected galaxies at similar redshifts (Madau et al. 1996). At present
there are large uncertainties in the LALA numbers due to the small
sample of spectroscopic confirmations.

While the original predictions were for strong Ly-$\alpha$ emission,
the recent detections are roughly a hundred times fainter.  Prior to
the discoveries of Ly-$\alpha$ emitters, many scenarios had been
proposed to explain the non-detections.  Now that we find Ly-$\alpha$
sources, understanding the faintness of this line will be
valuable to understanding star and galaxy formation this early in the
universe.

(A) It is possible that early star formation occurred in smaller,
sub-galactic units, which later merged to form the large galaxies we
see today.  In this, case the line luminosity L$_{Ly-\alpha}$ should
be small while its equivalent width remains large. Two of the
confirmed $z=4.5$ Ly-$\alpha$ sources lie only $5''$ apart and may
indeed be interacting or merging.

(B) Even modest amounts of dust can strongly quench the line:
Ly-$\alpha$ photons are resonantly scattered by neutral hydrogen,
which can force them to escape from galaxies by a random walk process
in which they traverse a much longer path length for dust absorption
than do continuum photons at similar wavelengths.  In this case, both
 L$_{Ly-\alpha}$ and $\hbox{EW}_{\hbox{\scriptsize Ly}\alpha}$ should be substantially reduced from the
simplest dust-free predictions. Most of the spectroscopically
confirmed LALA sources have rest-frame ${\hbox{EW}_{\hbox{\scriptsize Ly}\alpha}} \approx 100 \AA$ which
argues for dust-free star-formation scenarios. This by itself is not a
strong argument, since the narrow band selection picks out high
equivalent width sources.  Comparison of the populations of
line-selected and continuum selected sources would tell us how many
galaxies at these redshifts are dust-poor enough to show high equivalent width 
Ly-$\alpha$ emission.

(C) Similarly, early star-formation could have a longer time-scale
than a dynamical time postulated by PP67.  NIR colors along with stellar
population modelling can help to determine whether there is some relatively
old population (age $ > 10^7$ years), along with the young stars. 

(D) Even if the star-formation is dust-free, it is quite likely that
neutral gas extends beyond the star-formation regions. In this case
Ly-$\alpha$ photons will diffuse out of the surface area of the
larger HI envelope and Ly-$\alpha$ emission will be extended and 
fall below the surface brightness detection limits.

(E) Winds: In low redshift universe, the galaxies that show
Ly-$\alpha$ emission are not the most metal poor (and hence dust
free) galaxies, but galaxies which show outflows (Kunth et
al. 1998). Large velocity gradients doppler shift the Ly-$\alpha$
photons so they no longer resonantly scatter.  However, the high
redshift Ly-$\alpha$ sources emit $\sim 100$ times more in line
luminosity than these local sources, and it is not clear if the same
mechanism is responsible for the Ly-$\alpha$ escape in both
contexts.  If it is indeed winds that are responsible for the escape
of Ly-$\alpha$ photons at higher redshifts, these winds probably
have 100 times the kinetic energy of their local counterparts and
therefore are significant polluters of the intergalactic
medium. Either high resolution spectroscopy or morphological
comparison in continuum and Ly-$\alpha$ line emission will allow us
to search for evidence for such winds, and wind blown bubbles.

\section{Questions}
{\bf Avery Meiksin} Can you image these systems to see their morphologies, or can you only see the Ly-a emitting region? \\
Answer: These objects are at best marginally resolved from the ground; some
have little or no continuum detected. HST imaging in the Ly-$\alpha$ line and in the continuum would be very
interesting.
\\
\\
{\bf Cong Xu} Can you determine whether a source is an AGN?\\
Answer: We do not see broad emission lines. We cannot rule out weak,
narrow lined AGNs. To determine that we need spectra with enough S/N
to detect CIV at 1549 \AA\ or high resolution imaging capable of
revealing a point source. A hidden AGN would be best revealed in X-ray
imaging.


\begin{references}
{\small
\reference{} Cowie, L. L., \& Hu, E. M. 1998, \aj\ 115, 1319
	% blank field search at z=3.4, HDF and SSA22.
\reference{} Dey, A., Spinrad, H., Stern, D., Graham, J. R., \&
	Chaffee, F. H. 1998, \apjl\ 498, 93
	% z=5.34 galaxy (first one at z>5).
\reference{} Hu, E. M., Cowie, L. L., \& McMahon, R. G. 1998,
	\apjl\ 502, L99
	% The Density of Lyman Alpha Emitters at Very High Redshift
\reference{} Hu, E. M., McMahon, R. G., \& Cowie, L. L. 1999,
	\apjl\ 522, L9  % an extremely luminous z=5.74 galaxy
\reference{} Jannuzi, B. T., \& Dey, A., 1999, in ``Photometric Redshifts
    and High Redshift Galaxies'', ASP Conference Series, Vol. 191,
    editors R. J. Weymann, L. J. Storrie-Lombardi, M. Sawicki,
    and R. J. Brunner, p.111
\reference{} Kudritzki, R.-P.,  et al.2000, \apj\ 536, 19
	%  z=3.1 emitters from Virgo PNe surveys
\reference{} Kunth, D., Mas-Hesse, J. M., Terlevich, E., Terlevich, R., 
	Lequeux, J., \& Fall, S. M. 1998, \aap\ 334, 11
	% HST obs of local Lyman alpha emitters, and effects of outflows.
\reference{} Lowenthal, J. D., Koo, D. C., Guzman, R., Gallego, J.,
	Phillips, A. C., Faber, S. M., Vogt, N. P., Illingworth, G. D.,
	\& Gronwall, C. 1997, \apj\ 481, 673
	% Spectroscopy of z ~ 3 galaxies in the HDF
\reference{} Madau, P., Fergueson, H., Dickinson, M., Giavalisco, M., Steidel, C.C., Fruchter, A. 1996, MNRAS, 283, 1388.
\reference{} Oke, J.B. et al.~1995, PASP, 107, 375
\reference{} Partridge, R. B., \& Peebles, P. J. E. 1967, \apj\ 167, 868
	% Early prediction of Lyman alpha emission from protogalaxies
\reference{} Pritchet, C. J. 1994, \pasp\ 106, 1052
	% Big review on "The Search for Primeval Galaxies"
\reference{} Rhoads, J., Malhotra, S., Dey, A., Stern, D., Spinrad, H., Jannuzi, B. 2000, submitted to ApJL (Astro-ph/0003465).
\reference{} Steidel, C., Adelberger, K.L., Giavalisco, M., Dickinson, M., Pettini, M. 1999, \apj\ 519, 1
\reference{} Stern, D., \& Spinrad, H. 1999, \pasp\ 111, 1475
}
\end{references}
\end{document}